\documentclass{aastex}
\usepackage{emulateapj5}
\usepackage{apjfonts}
\usepackage{epsf}

\slugcomment{RESCEU-13/02 UTAP-420}
\shorttitle{SUPERNOVAE TIME DELAY STATISTICS}
\shortauthors{OGURI, SUTO, \& TURNER}
\begin{document}
\title{Gravitational Lensing Magnification and Time Delay Statistics
for Distant Supernovae}
%
\author{Masamune Oguri,\altaffilmark{1}
Yasushi Suto,\altaffilmark{1,2} and Edwin L. Turner\altaffilmark{3}}

\email{oguri@utap.phys.s.u-tokyo.ac.jp, suto@phys.s.u-tokyo.ac.jp,
 elt@astro.princeton.edu}

\altaffiltext{1}{Department of Physics, School of Science, University of
Tokyo, Tokyo 113-0033, Japan}
\altaffiltext{2}{Research Center for the Early Universe (RESCEU), School
of Science, University of Tokyo, Tokyo 113-0033, Japan}
\altaffiltext{3}{Princeton University Observatory, Peyton Hall,
Princeton, NJ 08544}
%
\received{2002 August 26}
\accepted{2002 October 3}
\begin{abstract}
Strong gravitational lensing of distant supernovae (SNe), particularly Type
Ia's, has some exploitable properties not available when other sorts of
cosmologically distant sources are lensed.  One such property is that
the ``standard candle'' nature of SN at peak brightness allows a
direct determination of the lensing magnification factor for each well
observed image.  Another is that the duration of a SN event is of the
same order as the differential time delays between the various lens
images for roughly galaxy mass lensing objects.  A relatively precise
constraint on each image's magnification leads to better constraints on
the lens mass model than are available in more familiar lens systems,
and the comparable time scales of the photometric event and the time
delay invite a variety of applications, including high precision
measurements of the delay and the targeting of especially interesting
phases of the explosion (including its very early stages) for intensive
observation when they appear in trailing images.

As an initial exploration of these possibilities we present calculations
of SN lensing statistics in a ``concordance cosmology'' assuming a simple
spherical model for lens mass distributions.  We emphasize magnification
and time delay effects.  Plausible SN surveys, such as the proposed {\it
SNAP} space mission, would discover several to some tens of strongly
lensed SNe Ia per year, and at least a few of these will be at redshifts
well beyond those that would be accessible via unlensed events. The
total number of such anomalously high redshift SNe Ia will be a useful
test of high redshift star formation models.  SN surveys of finite
duration will, of course, miss the appearance of some images, and the
effect becomes large when the delay approaches the survey duration; we
quantify this selection bias.  Finally, we investigate how well the
appearance of trailing images can be predicted based on various amounts
of available information on the lensing event.  Knowledge of the
magnification factor for the leading (and brighter) image makes it
possible to predict the appearance of a trailing image relatively
accurately if the lens redshift is also known.
\end{abstract}
%
\section{Introduction}

Recent systematic surveys of distant type Ia supernovae (hereafter SNe
Ia) strongly suggest the presence of {\it dark energy} which may
dominate the total energy density of the universe
\citep{riess98,perlmutter99}.  The reason that this rather surprising
conclusion is taken so seriously stems from the fact that the SNe Ia are
excellent (albeit empirical) ``standard candle'' distance indicators,
after an appropriate correction for the peak luminosity dependence on
the shape of the individual lightcurve \citep*{phillips93,riess96}.  The
exploitation of SNe Ia as cosmological probes has already been extensive
\citep{riess98,perlmutter99}, and plans for yet more extensive
observational studies, such as {\it
LSST}\footnote{http://www.lssto.org/lssto/} (Large-aperture Synoptic
Survey Telescope), are being rapidly developed.  The most ambitious
of these is the proposed satellite {\it
SNAP}\footnote{http://snap.lbl.gov/} (SuperNova/Acceleration Probe)
which would gather $\sim 2000$ SNe Ia per year by frequently imaging
$\sim 20$ square degrees of sky.

In this paper we explore some other ways in which the unique transient
``standard candle'' properties of SNe Ia might be exploited for those
few which happen to be strongly gravitationally lensed by an intervening
object of roughly galactic mass.  For the sake of specificity we will
adopt the observational parameters associated with the proposed {\it
SNAP} mission unless otherwise stated.

Very roughly 0.1 percent of sources at $z>1$ are expected to have
multiple images due to strong gravitational lensing
\citep*[e.g.,][]{turner84}, and {\it SNAP} therefore would find at least
$\sim 2$ lensed SNe Ia per year \citep[e.g.,][]{holz01}. They will be
qualitatively different from the lensed systems so far detected (e.g.,
$\sim 60$ QSO multiple-image systems\footnote{A summary of known lensed
QSO systems is available on the CASTLES homepage (Kochanek, C.~S.,
Falco, E.~E., Impey, C., Lehar, J., McLeod, B., \& Rix, H.-W.,
http://cfa-www.harvard.edu/castles/)}).  First of all, the time-delay
between the multiple images could be robustly and accurately determined
for each object, at least in principle.  Indeed, for a typical
cosmological lensing time delay, of an order of a year, it is unlikely
that one will observe all of the multiple images {\it simultaneously}
due to the finite duration of SNe Ia ($\sim$ a month). This also means
that some lensing events may be missed because the observation time is
finite; one or more multiple images may appear only before or after the
observing season or program.  This is in marked contrast to QSO
multiple-image systems where the images are observed simultaneously and
their presence and geometrical arrangement in the sky is often the chief
indication of lensing.  Furthermore, since the SNe Ia are believed to be
a reliable ``standard candle'', the magnification factor of the lensing
can be determined directly.  This is not feasible for any other sources
because an object's intrinsic luminosity is basically unknown.  This can
provide valuable additional information on the lensing potential.  With
such information, one may indeed {\it predict} the location and the
epoch of the additional trailing images, at least statistically.

In the present paper, we first analytically calculate the expected
number of lensed SNe Ia to be found by a {\it SNAP}-like survey,
including the effects of time delay bias due to a finite
observation/survey duration. Since lens systems with wider separations
have larger time delays on average, time delay bias becomes more
significant for bigger image separations.  Next we describe ways to
predict the appearance of additional images of lensed SNe Ia.  We then
briefly consider the consequences of using a different and perhaps more
realistic model for lens mass distributions.  In conclusion, we discuss
some of the implications and possible extensions of this work.

Since we are not here concerned with determining cosmological parameters
via lensing (although this constitutes one of the primary purposes of
the SNe Ia survey), unless otherwise specified we simply adopt the
lambda-dominated universe with $(\Omega_0, \lambda_0,h)=(0.3,0.7,0.7)$,
where $\Omega_0$ is the density parameter, $\lambda_0$ is the
cosmological constant, $h$ is the Hubble constant in units of $100{\rm
km\,s^{-1}Mpc^{-1}}$, the so called concordance cosmology
\citep{ostriker95,bahcall99}.

\section{Number Counts of Supernova Ia}\label{sec:snrate}

The number of SNe is closely related to the star formation rate $R_{\rm
SF}$ in the universe. For a Salpeter initial mass function ($\phi_{\rm
IMF}(M)\propto M^{-2.35}$, $0.1M_\odot<M<125 M_\odot$), the rate of Type
Ia events $R_{\rm SNeIa}$ is calculated as follows \citep*{madau98}
\begin{eqnarray}
 R_{\rm SNeIa}(z)=&&\nonumber\\
&&\hspace*{-16mm}\frac{\eta\int_0^{t(z)}R_{\rm SF}(z(t'))dt'
\int_{M_{\rm c}}^{8M_\odot}
\exp\left[-\left(t-t'-t_M\right)/\tau\right]
\phi_{\rm IMF}(M)dM}{\tau\int M\phi_{\rm IMF}(M)dM} ,
\label{snrate}
\end{eqnarray}
where $M_{\rm c}=\max[3M_\odot,(10{\rm Gyr}/t')^{0.4}M_\odot]$ is the
minimum mass of a star that reaches the white dwarf phase at time $t'$
(assuming all systems with the primary star of mass $3M_\odot \le M \le
8M_\odot$ are possible progenitors of SNe Ia), $t_M=10{\rm
Gyr}/(M[M_\odot])^{2.5}$ is the standard lifetime of a star of mass $M$,
$\tau$ is a characteristic explosion time scale (corresponding to an
effective ``delay'' time scale after the collapse of the primary star to
a white dwarf), $\eta$ is the SNe Ia explosion efficiency. We adopt
three representative star formation rates used by \citet{porciani01}.
The explicit forms of these in an Einstein-de Sitter universe are
\begin{eqnarray}
R_{\rm SF1}(z)&=&0.462h
\frac{\exp(3.4z)}{\exp(3.8z)+45}M_\odot{\rm yr^{-1}Mpc^{-3}},\\
R_{\rm SF2}(z)&=&0.230h
\frac{\exp(3.4z)}{\exp(3.4z)+22}M_\odot{\rm yr^{-1}Mpc^{-3}},\\
R_{\rm SF3}(z)&=&0.308h
\frac{\exp(3.05z-0.4)}{\exp(2.93z)+15}M_\odot{\rm yr^{-1}Mpc^{-3}}.
\end{eqnarray}
The first model (SF1) includes a correction for dust reddening, and
matches most measured UV-continuum and H$\alpha$ luminosity densities
\citep{madau00}. The second model (SF2) is also possible because of the
uncertainties associated with the incompleteness of data sets and the
amount of dust extinction \citep{steidel99}. The third model (SF3) is
considered because it has been suggested that the rates at high $z$ may
have been severely underestimated due to an unexpectedly large amount of
dust extinction \citep[e.g.,][]{blain99}. These star formation rates are
easily converted to those in different cosmologies \citep{porciani01},
and are considered to span the range of reasonably realistic
possibilities.

The number rate of SNe Ia which occur between $z$ and $z+dz$ is then
\begin{equation}
 \frac{dN}{dz}=
\frac{R_{\rm SNeIa}(z)}{1+z}
\frac{\Omega_{\rm A}D_{\rm A}^2(z)}{H(z)(1+z)}(1+z)^3,
\label{obsnum}
\end{equation}
where $\Omega_{\rm A}$ is the solid angle of the observed region and
$D_{\rm A}(z)$ is the angular diameter distance.

\section{Lensing Statistics}

\subsection{Image Separation and Time Delay Probability Distribution}

In this discussion, we omit the magnification bias
\citep{turner80,turner84} because the intrinsic luminosity function at
peak brightness of Type Ia SNe is quite narrow\footnote{This is
equivalent to assuming that the survey monitors every field sufficiently
frequently to catch each SN event near its peak brightness, as is the
case for {\it SNAP}.  The bias could not be neglected for a more
traditional survey which finds most objects well after their maximum
brightness.}.  We consider that magnification of SNe at redshifts beyond
those normally accessible to the survey, another form of magnification
bias, separately in the next subsection.

We now consider lensing objects with a the Singular Isothermal Sphere
(SIS) density profile:
\begin{equation}
 \rho(r)=\frac{v^2}{2\pi Gr^2},
\label{sis}
\end{equation}
where $v$ is a one-dimensional velocity dispersion and define the
characteristic scale length
\begin{equation}
 \xi_0=4\pi\left(\frac{v}{c}\right)^2\frac{D_{\rm OL}D_{\rm LS}}{D_{\rm OS}},
\end{equation}
where $D_{\rm OL}$, $D_{\rm OS}$, and $D_{\rm LS}$ are the angular
diameter distances to the lens, to the source, and between the lens and
source, respectively.  Then the lens equation becomes $y=x-x/|x|$, where
$x$ and $y$ are image and source positions in each plane normalized by
$\xi_0$ and $\xi_0D_{\rm OS}/D_{\rm OL}$, respectively. This has two
solutions $x_\pm=y\pm1$, and each image will be magnified by a factor
$\mu_\pm=(1/y)\pm1$.  The angular separation of two images is given by
\begin{equation}
 \theta=\frac{\xi_0(x_+-x_-)}{D_{\rm OL}}
=8\pi\left(\frac{v}{c}\right)^2\frac{D_{\rm LS}}{D_{\rm OS}}.
\end{equation}
The differential time delay between two images can also be calculated as
\begin{equation}
 c\Delta t(y)=32\pi^2\left(\frac{v}{c}\right)^4
\frac{D_{\rm OL}D_{\rm LS}}{D_{\rm OS}}(1+z_{\rm L})y.
\end{equation}

\begin{figure*}[t]
 \epsscale{1.4}
\plotone{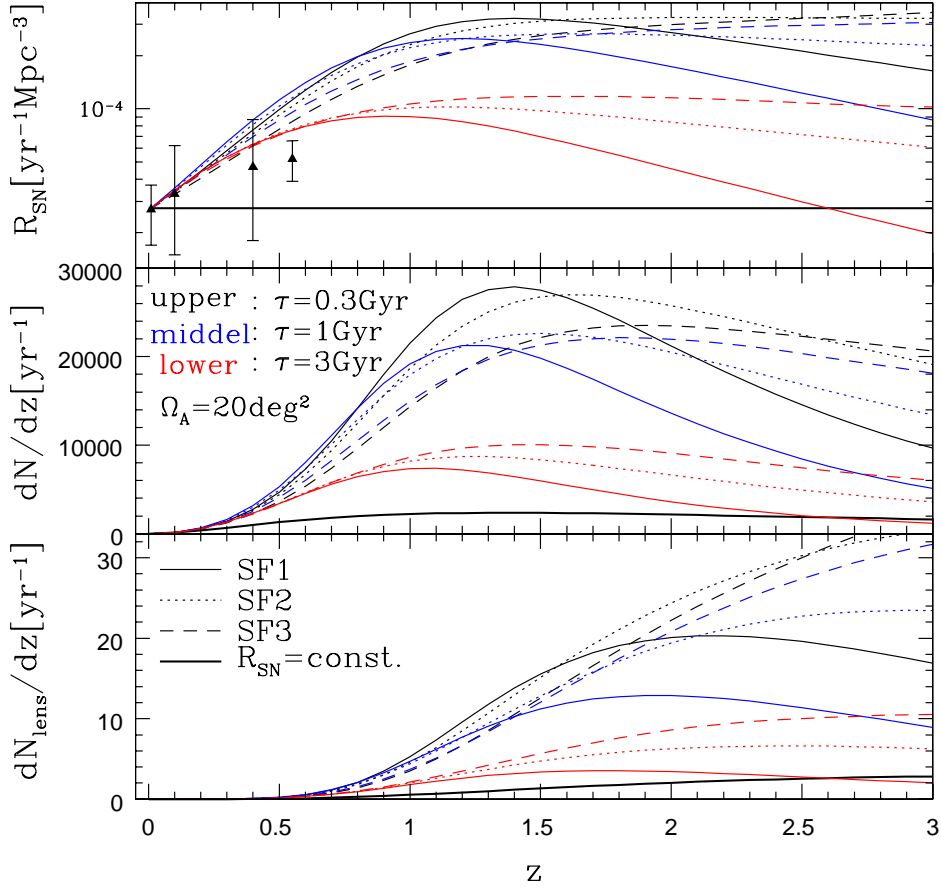} 
\caption{Supernova rates $R_{\rm SN}$ (eq. [\ref{snrate}]), number rates
 of SNe Ia $dN/dz$ (eq. [\ref{obsnum}]), and number rates of lensed SNIa
 $dN_{\rm lens}/dz=\int dN/dz P(z)dz$ as a function of redshift of SNe
 Ia, on the basis of the model described in \S \ref{sec:snrate}.  In the
 middle and bottom panels, observations of 20 square degrees field is
 assumed and no magnitude limit is imposed. For $R_{\rm SN}$, four
 observed SN Ia rates are also plotted by filled triangles; $z\sim 0.01$
\citep*{cappellaro99}, $z\sim 0.1$ \citep{hardin00}, $z\sim 0.4$
\citep{pain96}, and $z\sim 0.55$ \citep{pain02}. The efficiency
 parameter $\eta$ is adjusted so as to reproduce the observed SN Ia rate
 at $z\sim0.01$.  The ``concordance'' lambda-dominated universe with
$(\Omega_0, \lambda_0,h)=(0.3,0.7,0.7)$ is assumed. \label{fig:num}} 
\end{figure*}

The cumulative and differential probability distributions of strong
lensing are
\begin{equation}
 P(>\theta; z_{\rm S})
=\int_0^{z_{\rm S}}dz_{\rm L}
\int^\infty_{v_{\rm min}}\sigma_{\rm SIS}
\frac{c\,dt}{dz_{\rm L}}(1+z_{\rm L})^3\phi(v),
\end{equation}
\begin{equation}
\label{eq:condpnotb}
 P(\theta; z_{\rm S})=-\frac{d}{d\theta}P(>\theta; z_{\rm S}),
\end{equation}
where $\sigma_{\rm SIS}=\pi\xi_0^2$ is the cross section of strong
lensing, $v_{\rm min}=(\theta/8\pi)^{1/2}(D_{\rm OS}/D_{\rm
LS})^{1/2}c$, and $\phi(v)$ is the velocity function of lens galaxies.
We adopt the velocity function:
\begin{equation}
 \phi(v)dv=\frac{\Psi_*}{\ln 10}
\left(\frac{\sqrt{2}v}{v_*}\right)^\beta
\exp\left[-\left(\frac{\sqrt{2}v}{v_*}\right)^{\beta/2.5}\right]
\frac{dv}{v_*},
\label{velfunc}
\end{equation}
where $\Psi_*=7.3\times 10^{-2}h^3{\rm Mpc^{-3}}$, $\beta=-1.3$, and
$v_*=247{\rm km\,s^{-1}}$.  This distribution function is based on the
Southern Sky Redshift Survey and is derived by \citet{gonzalez00}.
Averaging the above probability function for fixed $z_{\rm S}$ over the
observed rate of SNe (eq.  [\ref{obsnum}]) yields the number rate of
strong lensing:
\begin{equation}
 P(\theta)=\int P(\theta; z_{\rm S})\frac{dN}{dz_{\rm S}}dz_{\rm S}.
\end{equation}

Similarly, the joint probability distributions of time delays and image
separations are (see eqs. [29]-[31] of Oguri et al. 2002)
\begin{eqnarray}
 P(>\Delta t, \theta; z_{\rm S})&&\nonumber\\
&&\hspace*{-15mm}=\int_0^{z_{\rm S}}dz_{\rm L}
\left[\frac{dv}{d\theta}\sigma_{\rm SIS}
N^T(>\Delta t)\frac{c\,dt}{dz_{\rm L}}(1+z_{\rm L})^3
\phi(v)\right]_{v=v_{\rm min}},
\end{eqnarray}
\begin{equation}
 P(\Delta t,\theta; z_{\rm S})
=-\frac{d}{d(\Delta t)}P(>\Delta t, \theta; z_{\rm S}),
\end{equation}
where $N^T(>\Delta t)=1-y_{\rm min}^2$ and $y_{\rm min}$ can be
calculated from $\Delta t=\Delta t(y_{\rm min})$. These joint
probability distributions divided by $P(\theta; z_{\rm S})$ give the
conditional probability distributions $P(>\Delta t|\theta; z_{\rm S})$
and $P(\Delta t|\theta; z_{\rm S})$.

Figure \ref{fig:num} shows the predicted number rates of SNe Ia and also
the expected number rate of lensed SNe Ia. The number rates of SNe Ia
are calculated from equation (\ref{obsnum}). It is clear from this
figure that the number of SNe Ia strongly depends on the star formation
rate and its evolution, as indicated by \citet{madau98}. The number of
lensed SNe Ia also shows large differences between models, reflecting the
above sensitive model-dependence of the number of SNe Ia. Our
calculation predicts that the number of strongly lensed SNe Ia will be
between a few and a few tens per year for a {\it SNAP}-like survey. These
lensing rates are roughly consistent with those calculated by
\citet{wang00} and \citet{holz01}.

\subsection{SNe beyond the Magnitude Limit}

Gravitational lensing magnifies SNe, thus some SNe which exceed the
magnitude limit if they are unlensed might be observed due to their
magnification \citep[e.g.,][]{kolatt98,porciani00,sullivan00,goobar02}.
If we neglect the effect of gravitational lensing, the apparent
magnitude of SNe at peak can be expressed as
\begin{equation}
 m_X=M_B+5\log (D_{\rm L}(z)[{\rm Mpc}])+25+K_{BX},
\end{equation}
where $m_B=-19.4$ is the peak magnitude of SNe Ia, $D_{\rm
L}(z)=(1+z)^2D_{\rm A}(z)$ is luminosity distance, and $K_{BX}$ is the
single- or cross-filter K-corrections. Therefore, even for high-$z$ SNe
Ia whose unlensed apparent magnitude exceeds the magnitude limit,
$m_X>m_{\rm lim}$, they are observed if magnified by a factor $\mu$
satisfying
\begin{equation}
 \mu\geq 10^{0.4(m_X-m_{\rm lim})}\equiv \mu_{\rm min}.
\end{equation}
We consider the following two cases: (1) $\mu_->\mu_{\rm min}$. This
corresponds to the case that both lensed images are observed. (2)
 $\mu_+>\mu_{\rm min}$. This means that at least one image is observed.
In each case, the cross section for lensing is
\begin{equation}
 \sigma(\mu_\mp>\mu_{\rm min})=\frac{\sigma_{\rm SIS}}{(\mu_{\rm min}\pm 1)^2}.
\end{equation}
We calculate the numbers of such lensed SNe Ia, and the results are
shown in Figure \ref{fig:num_hz}. We assume the magnitude limit is
$m_{\rm lim}=30$, and also impose a requirement that the photometry must
extend to 3.8 magnitudes below peak. We take account of the intrinsic
dispersion of SNe Ia peak magnitudes assuming the Gaussian
distribution with dispersion $\sigma_m=0.15$ \citep{porciani00}. As seen
in the figure, the expected number of such lensed SNe Ia depends very
strongly on models of star formation history. In most models, however,
more than one high-$z$ ($z\sim 3$) SNe Ia per year is expected to be
observed. Actually the difference simply comes from the difference in
supernova rates, thus we can infer supernova rates at high redshifts
from the observed number of lensed SNe Ia. The supernova rate as a
function of redshift is useful not only to constrain progenitor models
and star formation history \citep[e.g.,][]{yungelson00} as shown in
these plots, but also to test other possible redshift dependence of
the SNe Ia rate \citep{kobayashi98}. The SNe Ia rate also depends on the
assumed cosmological parameters. The cosmological parameters
$\Omega_0$ and $\lambda_0$ are now determined with $\sim 10$\% accuracy
in the combined analysis of Cosmic Microwave Background (CMB)
experiments, SNe Ia, and large scale structure surveys
\citep[e.g.,][]{sievers02}, and we assume the $\Omega_0$ uncertainty to be
$\Omega_0=0.3\pm 0.05$ in the flat universe ($\Omega_0+\lambda_0=1$).
We also plot the resulting numbers of lensed SNe Ia for the constant $R_{\rm
SN}$ model due to the $\Omega_0$ uncertainty. We find that this level of
uncertainty in $\Omega_0$ does not significantly change the expected number of
SNe so that we could still distinguish between models of the star formation
history.

\vspace{0.5cm}
\centerline{{\vbox{\epsfxsize=8.4cm\epsfbox{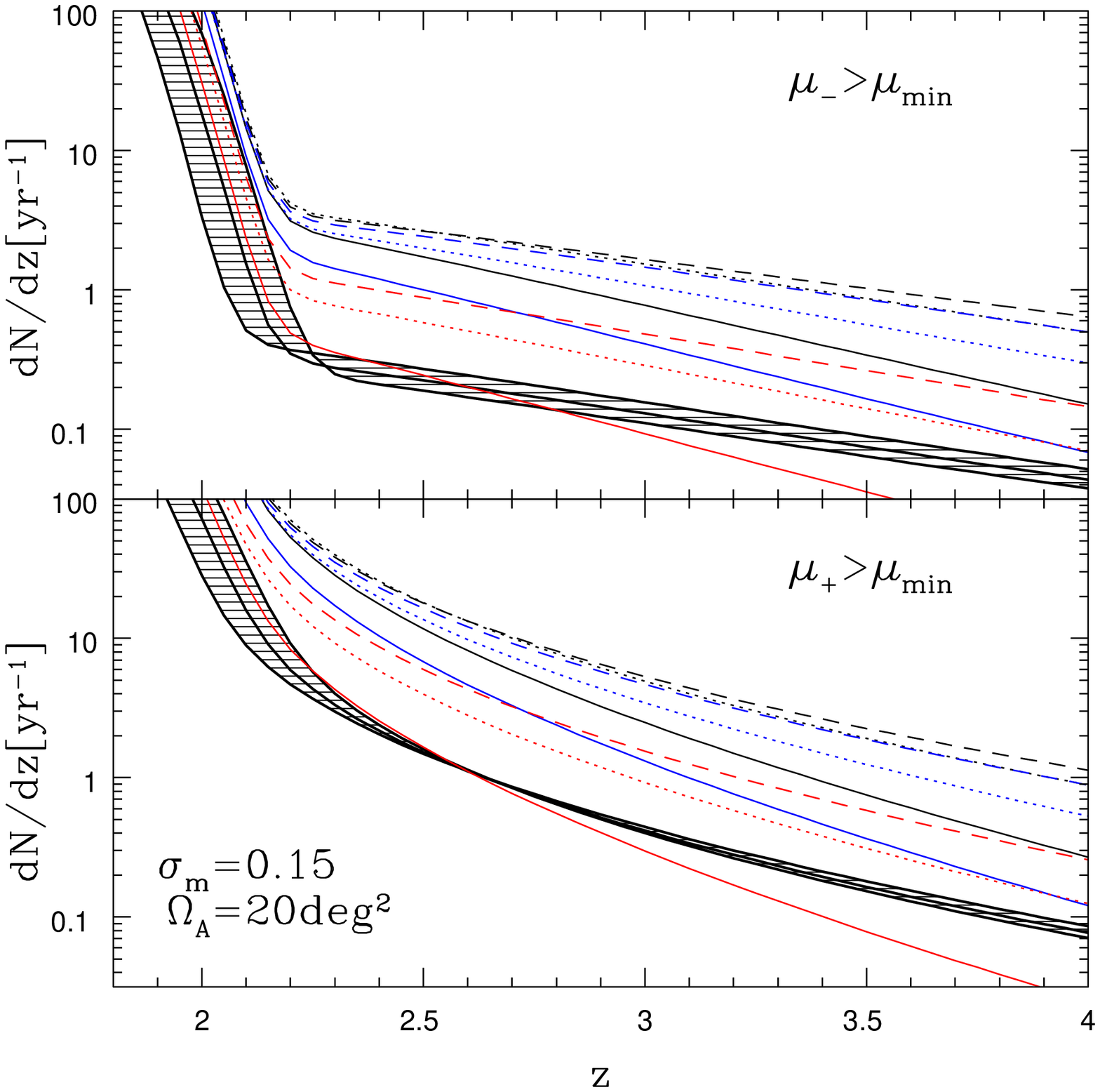}}}}
\figcaption{Numbers of lensed SNe Ia whose unlensed apparent magnitudes are
 fainter than the magnitude limit $m_{\rm lim}=30$ for the observed
 region of $\Omega_{\rm A} = 20$ deg$^2$.  Lines are same as in Figure
 \ref{fig:num}.  For the constant $R_{\rm SN}$ model, we show the effect
 of different cosmological parameters ($\Omega_0=0.3\pm 0.05$,
 $\Omega_0+\lambda_0=1$) by the same three lines and shadings. We also
 impose a requirement that the photometry must
 extend 3.8 magnitudes below peak  brightness (thus effectively
$m_{\rm lim}=26.2$), and K-corrections are neglected. The limiting peak
 magnitude $m_{\rm lim}=26.2$ roughly corresponds to $z\sim 1.7$.  The
 scatter of peak magnitude, $\sigma_m=0.15$, is taken into account. A
 lambda-dominated universe is again assumed. \label{fig:num_hz}}
\vspace{0.5cm}

If both images are observed, we can infer the intrinsic magnitude of
SNe Ia. For example, from the ratio of luminosities,
\begin{equation}
 r=\frac{\mu_+}{\mu_-},
\end{equation}
the magnification factors become
\begin{equation}
 \mu_+=\frac{2r}{r-1},
\end{equation}
\begin{equation}
 \mu_-=\frac{2}{r-1}.
\end{equation}
The magnification factors also can be derived from the image separation
and differential time delay if the redshift of the lens object is
measured (see eq. [\ref{predict_deltat}]). The reconstruction of
magnification factors could be used, in principle, to derive the
distance-redshift relation at high-$z$ and to estimate the cosmological
parameters more robustly.

\section{Time Delay Bias}\label{sec:tdb}

Since there is a time delay between multiple images, strong lensing
statistics of transient phenomena such as SNe inevitably involve some
missing events due to the finite duration of the survey observations.
Therefore, in strong lensing statistics of SNe we should take account of
this ``time delay bias'', especially for large image separations.

For the joint probability distributions, the time delay bias is included
as follows
\begin{equation}
 P^{\rm TB}(\Delta t, \theta; z_{\rm S})
=P(\Delta t, \theta; z_{\rm S})f(\Delta t),
\end{equation}
\begin{equation}
 P^{\rm TB}(>\Delta t, \theta; z_{\rm S})
=\int_{\Delta t}^\infty
P(\Delta t', \theta; z_{\rm S})f(\Delta t')d(\Delta t'),
\end{equation}
where $f(\Delta t)$ is the fraction of lenses with time delays $\Delta
t$ that can be observed (the superscript TB explicitly indicates the
Time-delay Bias). The image separation distribution then becomes
\begin{equation}
  P^{\rm TB}(\theta; z_{\rm S})
=\int_0^\infty P(\Delta t', \theta; z_{\rm S})f(\Delta t')d(\Delta t').
\end{equation}
Then the corresponding {\it conditional}
probability distributions are computed as
\begin{equation}
 P^{\rm TB}(\Delta t|\theta; z_{\rm S})
=\frac{P^{\rm TB}(\Delta t, \theta; z_{\rm S})}
{P^{\rm TB}(\theta; z_{\rm S})},
\end{equation}
using $P^{\rm TB}(\theta; z_{\rm S})$ instead of $P(\theta; z_{\rm S})$
which does not take account of the time-delay bias
(eq.[\ref{eq:condpnotb}]).

If the observational monitoring is carried out continuously for a
period of $t_{\rm obs}$, for instance, $f(\Delta t)$ is given by
\begin{eqnarray}
 f(\Delta t)&=&
\left\{
\begin{array}{@{\hspace{0.6mm}}ll}
\displaystyle{1-\frac{\Delta t}{t_{\rm obs}}} &
\mbox{$(\Delta t < t_{\rm obs})$},\\
\displaystyle{0} & \mbox{$(\Delta t > t_{\rm obs})$}.
\end{array}
\right.
\label{fdeltat}
\end{eqnarray}
Then $P^{\rm TB}(\theta; z_{\rm S})$ reduces to
\begin{eqnarray}
 P^{\rm TB}(\theta; z_{\rm S})&
=&\left[1-\frac{1}{t_{\rm obs}}
\int_0^{t_{\rm obs}}P(>\Delta t|\theta; z_{\rm S})d(\Delta t)\right]
P(\theta; z_{\rm S})\nonumber\\
&\equiv& T(\theta,z_{\rm S})P(\theta; z_{\rm S}).
\label{prob_tdb}
\end{eqnarray}
This means that the time delay bias for the image separation probability
distribution is simply expressed by the multiplication factor
$T(\theta,z_{\rm S})$.

\begin{figure*}[t]
\epsscale{1.4} 
\plotone{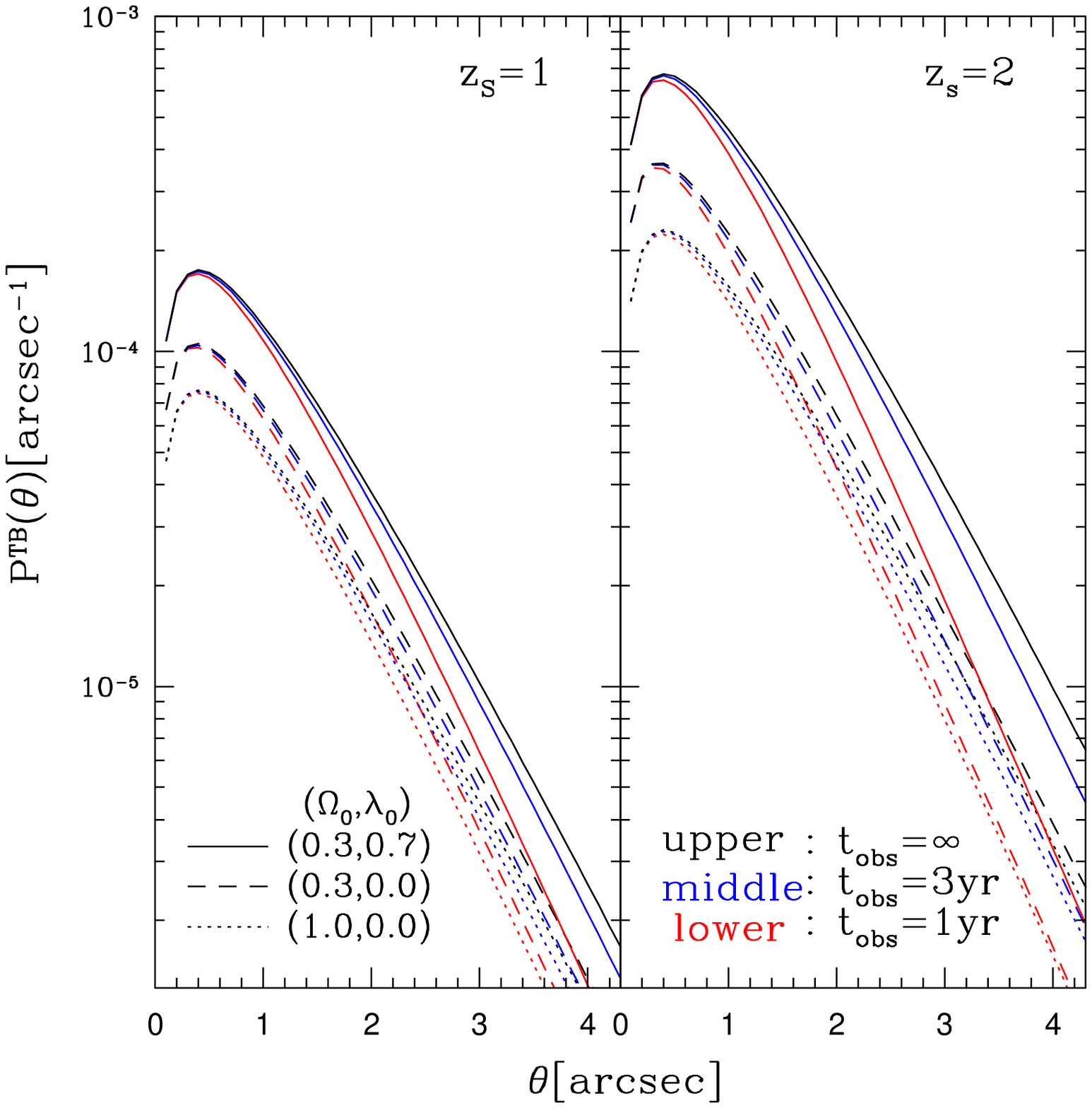} 
\caption{Probability distributions of strong gravitational lensing. The
 redshift of SNe is fixed to $z_{\rm S}=1$ ({\it left}) or $2$ ({\it
 right}). The time delay bias (\S \ref{sec:tdb}) is included, and
 continuous observations of $t_{\rm obs}=\infty$, $3{\rm yr}$, and
 $1{\rm yr}$ are considered. These probabilities are plotted for the
 three cosmological models with $h=0.7$;
 $(\Omega_0,\lambda_0)=(0.3,0.7)$, $(0.3,0.0)$, and $(1.0,0.0)$, in
 solid, dashed and dotted lines, respectively. \label{fig:prob}}  
\end{figure*}

Probability distributions of strong gravitational lensing including time
delay bias (eq. [\ref{prob_tdb}]) are shown in Figure \ref{fig:prob},
where we consider three cosmological models with $h=0.7$;
$(\Omega_0,\lambda_0)=(0.3,0.7)$, $(0.3,0.0)$, and $(1.0,0.0)$.  These
probability distributions have been used to constrain cosmological
constant \citep{turner90,fukugita92,kochanek96,chiba99}, and are indeed
useful as an independent measurement of cosmological parameters in the
case of SNe survey \citep{wang00,holz01,goobar02}.  These plots show that the
time delay bias is more important for larger $\theta$, because the time
delay $\Delta t$ is larger on average as $\theta$ increases. The amount
of the time delay bias is, however, quite small for lensing of typical
an image separation of $\theta\sim 1''$, if the observation time is
larger than 1 year. For larger separation lensing, $\theta\sim 3''$, the
time delay bias changes lensing probabilities by a factor 2 and thus is
important in quantitative discussions.

\begin{figure*}[t]
\epsscale{1.4}
\plotone{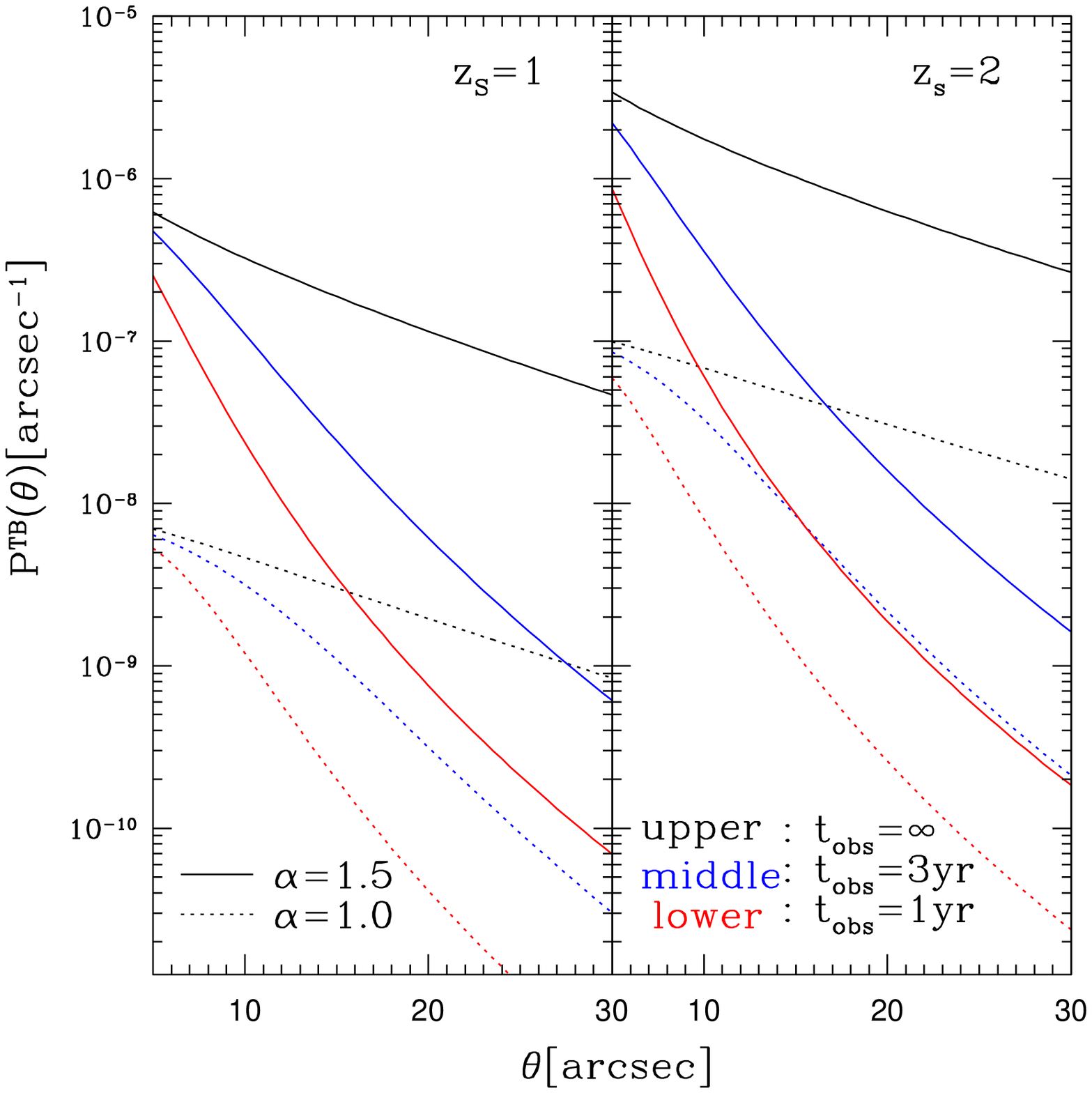}
\caption{Probability distributions of strong lensing at wide separation,
 calculated by assuming the generalized NFW density profile (eq.
 [\ref{nfw}]) and \citet{sheth99} mass function. We adopt the cold dark
 matter model in a lambda-dominated universe with the mass fluctuation
 amplitude $\sigma_8=1$. The time delay bias (\S \ref{sec:tdb}) is
 included. \label{fig:prob_widesep}}   
\end{figure*}

Since the time delay bias is more important for larger separation
lensing, we also calculate probability distributions for wide separation
lensing in the lambda-dominated Cold Dark Matter (CDM) model, assuming a
fluctuation amplitude $\sigma_8=1$ for simplicity.
Although the expected wide separation lensing rate due to CDM halos is
much smaller than for galaxy lensing, we still expect that wide
separation lensing will be observed because core-collapse SNe as well as
SNe Ia can be used for lensing statistics, which greatly increases the
number of SNe \citep[e.g.,][]{goobar02}. Wide separation lensing is
expected to reflect the properties of dark halos rather than (the
visible parts of) galaxies, thus it has been used to constrain the
abundance of dark halos \citep{kochanek95} and density profile of dark
halos \citep*{maoz97,wyithe01,keeton01b,takahashi01,li02,oguri02a}.  We
adopt the generalized form \citep{zhao96,jing00a} of the density profile
proposed by \citet*[][hereafter NFW]{navarro97}:
\begin{equation}
 \rho(r)=\frac{\rho_{\rm crit}\delta_{\rm c}}
{\left(r/r_{\rm s}\right)^\alpha\left(1+r/r_{\rm s}\right)^{3-\alpha}},
\label{nfw}
\end{equation}
where $r_{\rm s}=r_{\rm vir}/c_{\rm vir}$ and $c_{\rm vir}$ is the
concentration parameter. We adopt the mass and redshift dependence
reported by \citet{bullock01} for $\alpha=1$, and generalize it to
$\alpha\neq1$ by the multiplicative factor $(2-\alpha)$
\citep{keeton01b}. We also take account of scatter of the concentration
parameter which has a log-normal distribution with a dispersion of
$\sigma_{\rm c}=0.18$ \citep{jing00b,bullock01}.  The characteristic
density $\delta_{\rm c}$ can be computed using the spherical collapse
model \citep*[see][]{oguri01}.  As the mass function of dark halos, we
adopt the fitting form derived by \citet{sheth99}. From these, we
predict probability distributions of wide separation lensing taking
account of the time delay bias (see Oguri et al. 2002 for the
calculation of time delay probability distributions in the case of
generalized NFW density profile), and results are shown in Figure
\ref{fig:prob_widesep}. Here we focus on large separation lensing
($\theta>5''$) because the relation between small and large separation
lensing depends strongly on the model of galaxy formation and is
difficult to determine unambiguously \citep{oguri02b}.  This plot
indicates that time delay bias is quite significant; it can
suppress the lensing probability by one or two orders of magnitude. It
is also found that the suppression due to time delay bias is larger for
$\alpha=1.5$ than $\alpha=1.0$ because for a fixed separation $\theta$,
the time delays in the case of $\alpha=1.5$ are on average larger than
those in the case of $\alpha=1.0$ \citep{oguri02a}. This slightly
compensates for the difference of lensing probabilities between various
values of $\alpha$, but the difference is still large (about one order
of magnitude between $\alpha=1.5$ and $\alpha=1.0$). Therefore we
conclude that statistics of wide separation SNe lensing still could provide a
useful probe of density profile.

\begin{figure*}[t]
\epsscale{1.4}
\plotone{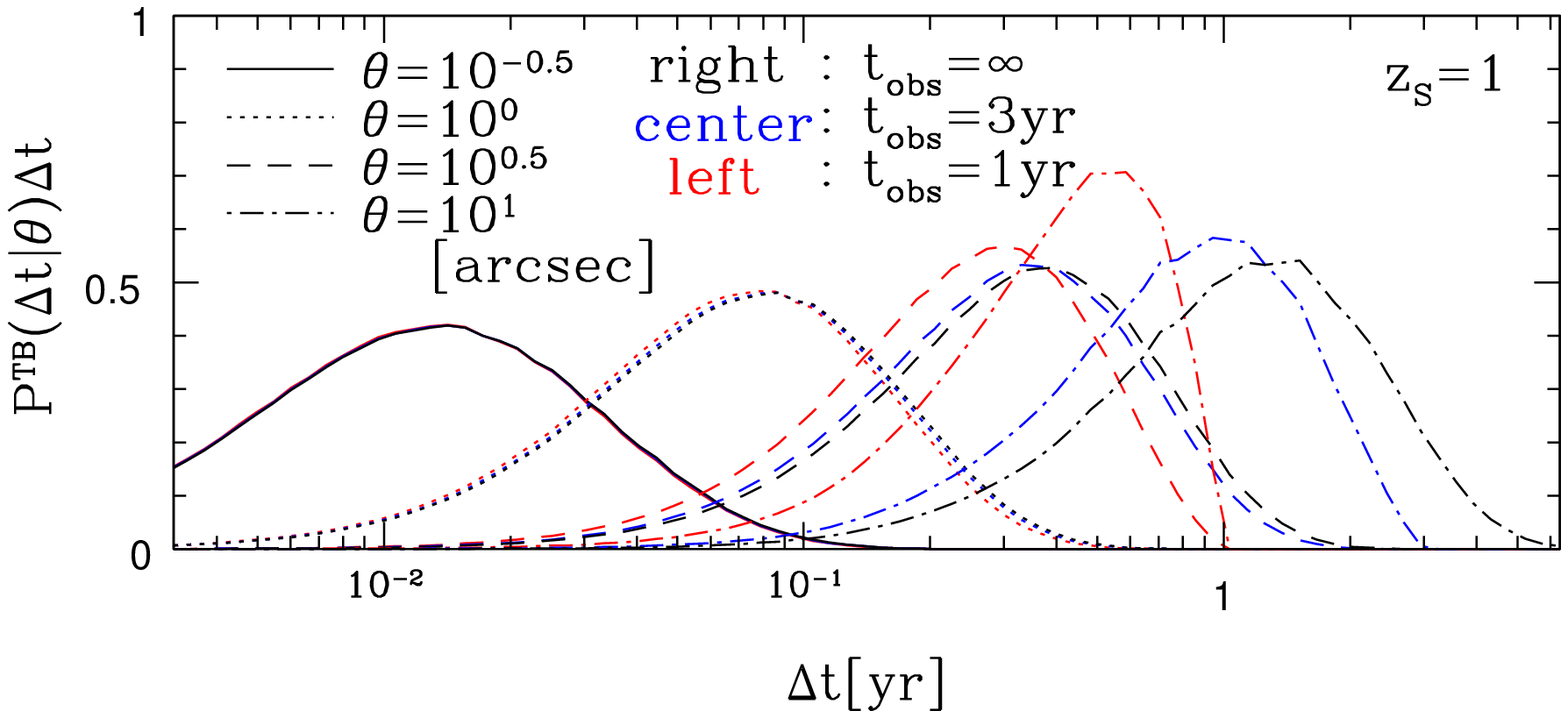}
\caption{Conditional probability distributions of differential time
 delays. The time delay bias (\S \ref{sec:tdb}) is included (as in  Fig.
 \ref{fig:prob}). A lambda-dominated universe is assumed.
 \label{fig:diff}} 
\end{figure*}

We also show the effect of the time delay bias on the time delay
probability distribution in Figure \ref{fig:diff}. As seen in the
figure, the time delay probability distributions for large $\theta$ are
strongly affected by the finite duration of observations.

\section{Predictions for Time-Delayed (Trailing) Images}

\begin{figure*}[t]
\epsscale{1.4}
\plotone{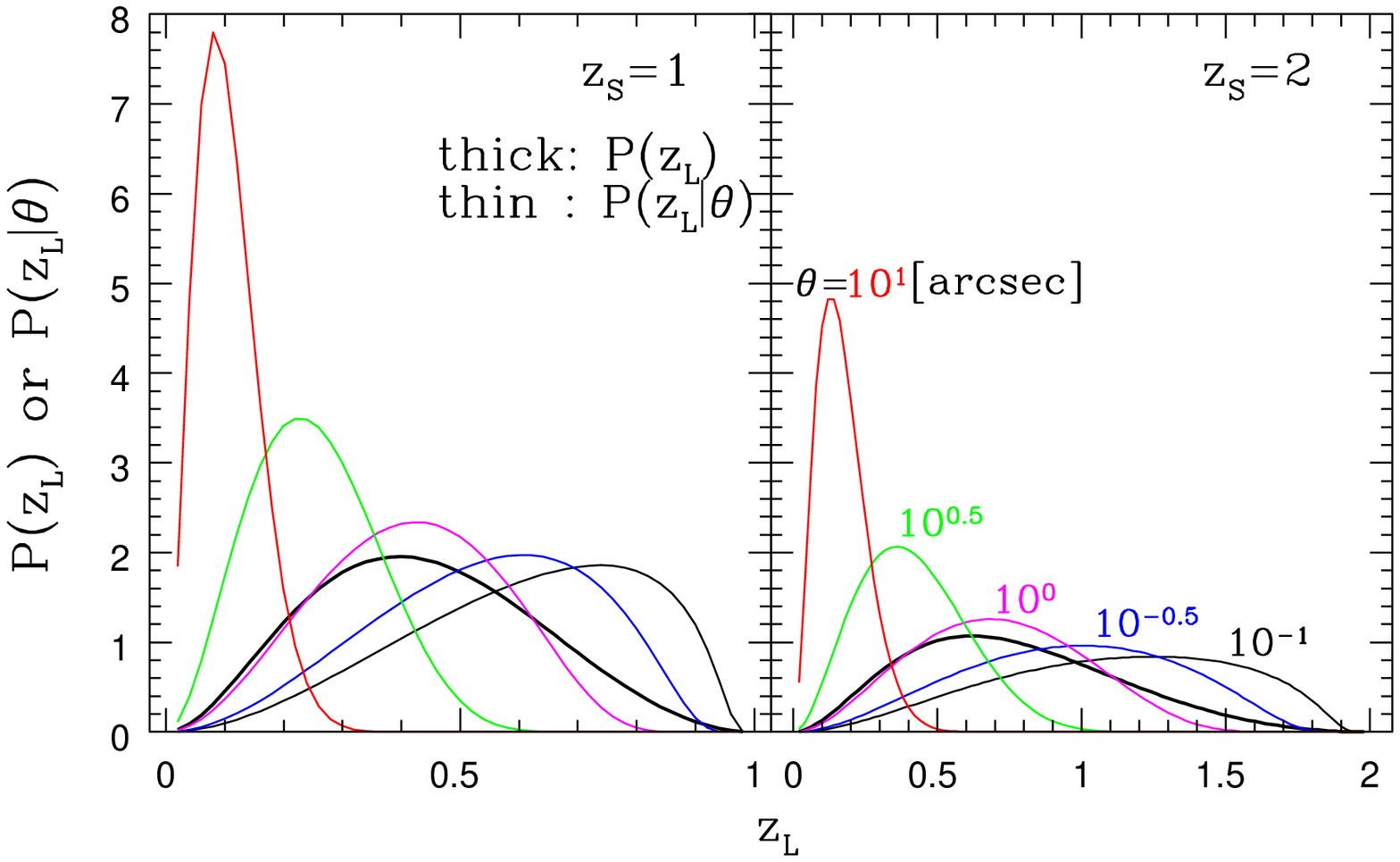}
\caption{Distributions of lens redshift $z_{\rm L}$ (eqs.
 [\ref{dpdz_theta_cond}] and [\ref{dpdz_notheta}]). Thick lines are
 probability distributions of $z_{\rm L}$ when $\theta$ is unrestricted,
 $P(z_{\rm L})$ (eq. [\ref{dpdz_theta_cond}]) while thin lines are those
 when $\theta$ is fixed, $P(z_{\rm L}|\theta)$ (eq.
 [\ref{dpdz_notheta}]). A lambda-dominated universe is assumed.
 \label{fig:dpdz}} 
\end{figure*}

Since the primary purpose of the proposed SNe Ia surveys is to construct
the Hubble diagram, one may always assume that the source redshift of
each SNe Ia, $z_s$, is determined spectroscopically. We are interested
in unusually bright SNe Ia in the diagram. If the lensing object is
approximated by an SIS, there are two images, $x_\pm$, and the brighter
image ($x_+$) arrives first. Thus the difference between the observed
magnitude of those outliers (with respect the typical magnitude of the
SNe Ia) and the average magnitude at the redshift can be ascribed to the
lensing magnification $\mu_+$.  While the standard deviation of the SNe
Ia peak magnitude corrected for the lightcurve shape method is typically
$0.15$mag \citep{porciani00}, multiple lensing images are produced if
$\mu_+ \ge 2$ or equivalently more than $0.75$mag. Thus 5$\sigma$
outliers are strong candidates for multiple image lensed SNe Ia.

The lensing object (most likely an early-type galaxy) for those
candidates is typically half way out in affine or angular diameter
distance.  For instance, an $L^*$ galaxy at $z=0.5$ is about $24$th
magnitude in the B-band, and the surface number density at this
magnitude limit is $\sim 30$ per arcmin$^2$ or so.  Since the typical
image separation $\theta_+$ is $1''$, the average number of galaxies
within that angular separation from each lensed SNe Ia is much less
than 1. Therefore it should be relatively easy to find candidates for
the lensing galaxy and thus determine $\theta_+$, and even possibly the
redshift of the candidate galaxy $z_{\rm L}$.

Our next goal is to predict the location of the second image $\theta_-$,
and the time delay $\Delta t$ by which the second image trails the first
and brighter one.  We consider the following three cases: (1) the
magnification factor $\mu_+$, the redshift of the lens object $z_{\rm
L}$, and the angular separation from the lens object
$\theta_+=\xi_0x_+/D_{\rm OL}$ are all known; (2) only $\mu_+$ and
$\theta_+$ are known; (3) just $\mu_+$ is known.  In doing so, we have
to assume a specific model for the lensing potential. We mainly show
results for the SIS lensing model (eq. [\ref{sis}]) but also present a
comparison with the NFW halo model (eq. [\ref{nfw}]).

\subsection{Case 1: $\mu_+$, $\theta_+$, and $z_{\rm L}$ are known}\label{sec:case1}
We can obtain these three quantities when the lens candidate is
identified and the redshift of that lens candidate is known.
In this case, both the separation $\theta$ and the time delay $\Delta
t$ are uniquely determined as
\begin{equation}
 \theta=2\theta_+\frac{\mu_+-1}{\mu_+} ,
\label{predict_sep}
\end{equation}
and
\begin{equation}
 \Delta t=\frac{1}{2c}\frac{D_{\rm OL}D_{\rm OS}}{D_{\rm LS}}
(1+z_{\rm L})\theta^2\frac{1}{\mu_+-1}.
\label{predict_deltat}
\end{equation}

We now consider errors induced by observable quantities. We assume that
the redshifts are measured spectroscopically, thus errors from redshifts
are negligible. As for the image separation $\theta_+$, SNAP would have
an angular resolution of $0.1''$, which is also sufficient to determine
$\theta_+$ accurately for most purposes.  The most important source of
uncertainty is therefore $\mu_+$, because the magnification estimates
may be inaccurate due to substructure in the lens galaxies
\citep{mao98}, dust extinction and/or the intrinsic spread in corrected
SNe Ia peak luminosities, photometry, K-corrections and so forth. If
$\mu_+$ is sufficiently larger than 1 (this is correct for most strong
lensing cases), we obtain $\Delta t\propto (\mu_+)^{-1}$ from equations
(\ref{predict_sep}) and (\ref{predict_deltat}). This means that errors
in $\mu_+$ directly affect $\Delta t$; e.g., a 20\% error in $\mu_+$
results in a 20\% uncertainty in the $\Delta t$ estimation.

\subsection{Case 2: $\mu_+$ and $\theta_+$ are known}\label{sec:case2}

\begin{figure*}[t]
\epsscale{1.8}
\plotone{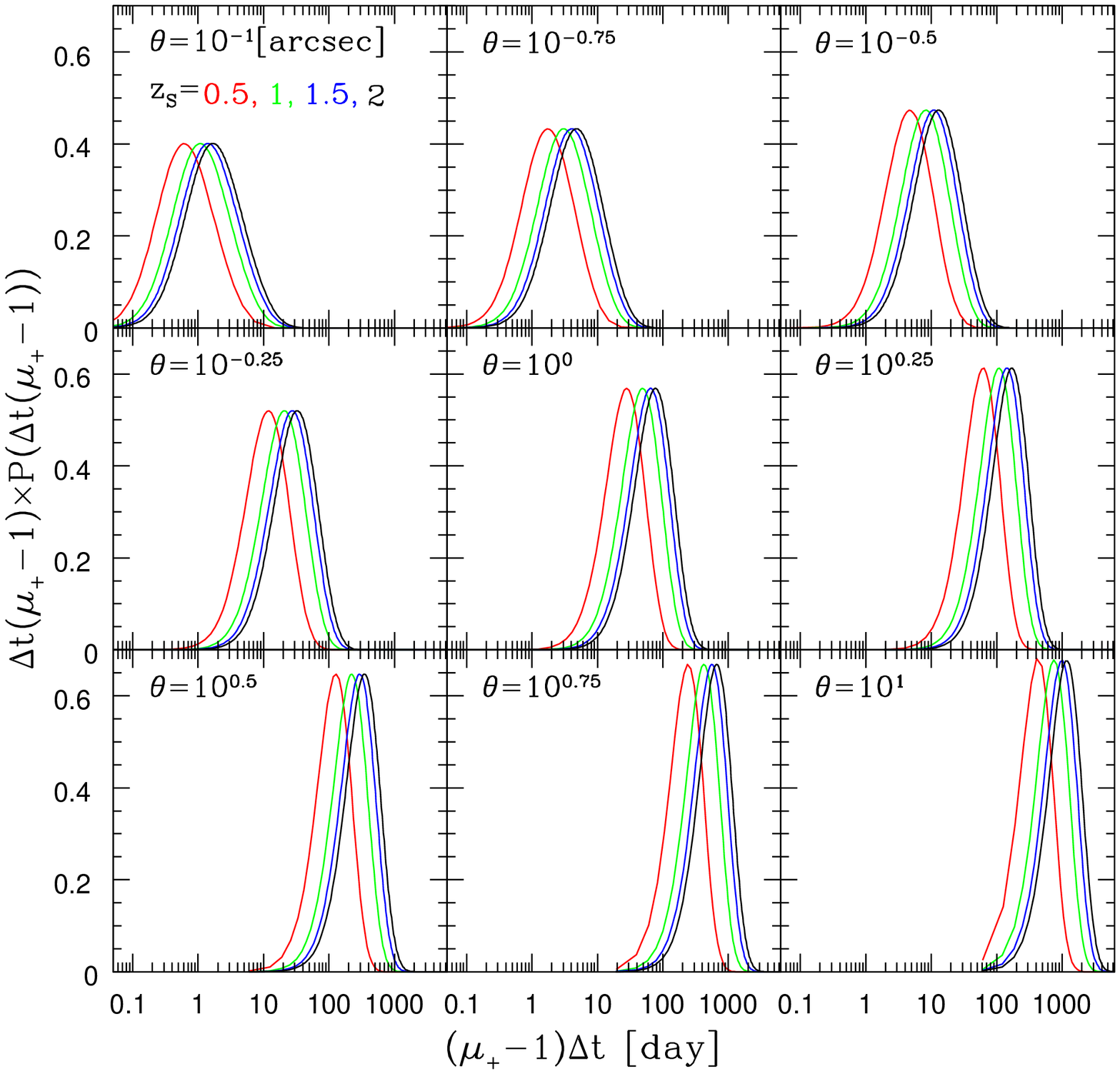}
\caption{Predictions for the trailing image in ``Case 2'' (\S
 \ref{sec:case2}). Probability distributions of $(\mu_+-1)\Delta t$ (eq.
 [\ref{dtmu1}] times $(\mu_+-1)\Delta t$) are plotted for various
 $z_{\rm S}$ and $\theta$. A lambda-dominated universe is assumed.
 \label{fig:predict_theta}}  
\end{figure*}

This is the case in which there is a lens candidate but the redshift is
not known (yet). We can predict the image separation from equation
(\ref{predict_sep}), but the time delay is ambiguous due to $z_{\rm L}$,
and has some probability distribution which reflects the distribution of
$z_{\rm L}$. We rewrite equation (\ref{predict_deltat}) as
\begin{equation}
 \frac{\Delta t(\mu_+-1)}{\theta^2}
=\frac{1}{2c}\frac{D_{\rm OL}D_{\rm OS}}{D_{\rm LS}}(1+z_{\rm L})\equiv F.
\end{equation}
Therefore we can obtain the probability distribution for the combination
of $\Delta t(\mu_+-1)$, instead of $\Delta t$ alone. To derive this, we
should calculate the conditional probability distribution of $z_{\rm L}$
for fixed $\theta$:
\begin{equation}
 P(z_{\rm L}|\theta; z_{\rm S})
=\frac{P(z_{\rm L},\theta; z_{\rm S})}{P(\theta; z_{\rm S})},
\label{dpdz_theta_cond}
\end{equation}
where
\begin{equation}
 P(z_{\rm L},\theta; z_{\rm S})
=\left[\frac{dv}{d\theta}\frac{d\phi}{dv}
\frac{c\,dt}{d_{\rm L}}(1+z_{\rm L})^3\sigma_{\rm SIS}\right]_{v=v_{\rm min}}.
\label{dpdz_theta}
\end{equation}
We plot $P(z_{\rm L}|\theta; z_{\rm S})$ for various $\theta$ values IN Figure
\ref{fig:dpdz}. Then the distribution of $\Delta t(\mu_+-1)$ is
\begin{equation}
 P(\Delta t(\mu_+-1))
=P(z_{\rm L}|\theta; z_{\rm S})
\left(\frac{dF}{dz_{\rm L}}\right)^{-1}\frac{1}{\theta^2}.
\label{dtmu1}
\end{equation}
The results are shown in Figure \ref{fig:predict_theta}. These results
are of course affected by observational uncertainties of $\mu_+$ as
mentioned in \S \ref{sec:case1}. If $\mu_+$ is sufficiently large,
errors in $\mu_+$ do not affect $\theta$ since $\theta \sim 2\theta_+$
(see eq. [\ref{predict_sep}]). On the other hand, $\Delta t$ is directly
affected by $\mu_+$ because we obtain the probability distribution for
$\Delta t(\mu_+-1)$. The width of this probability distribution is,
however, about one order of magnitude, while errors induced by $\mu_+$
are probably only a factor of 2 or so \citep{metcalf01,chiba02,dalal02}.
Therefore errors in $\mu_+$ are not so serious as in the previous case.

In practice, we find a fitting formula of $P(\Delta t(\mu_+-1))$:
\begin{eqnarray}
 P(\Delta t(\mu_+-1))d(\Delta t(\mu_+-1))=&&\nonumber\\
&&\hspace*{-43mm}\frac{1}{\sqrt{2\pi}\sigma}
\exp\left[-\frac{\left\{\ln\left(\Delta t(\mu_+-1)\right)-\ln a\right\}^2}
{2\sigma^2}\right]d\ln\left(\Delta t(\mu_+-1)\right),
\end{eqnarray}
\begin{equation}
 a=4.28(-110+158z_{\rm S}^{0.212})\theta^{1.63}(3.68+\theta)^{-0.952}
\,{\rm [day]},
\end{equation}
\begin{eqnarray}
 \sigma&=&0.693-0.115(\ln\theta)+0.0211(\ln\theta)^2\nonumber\\
&&+0.00347(\ln\theta)^3-0.00106(\ln\theta)^4,
\end{eqnarray}
where $\theta$ is in units of arcsec. This formula is valid for
$0.5\lesssim z_{\rm S}\lesssim 4.0$ and $0.1''\lesssim \theta \lesssim
10''$ which cover the range of our typical interest. The accuracy is
$\lesssim 10$\% around the peak (within $\sim 1.5\sigma$).

\subsection{Case 3: $\mu_+$ is known}\label{sec:case3}

This is the case in which a candidate for the lensing object is not
identified. The distribution of $z_{\rm L}$ for an unrestricted
$\theta$ is
\begin{equation}
 P(z_{\rm L}; z_{\rm S})=\frac{1}{P(z_{\rm S})}
\int_0^\infty dv\frac{d\phi}{dv}\frac{c\,dt}{dz_{\rm L}}
(1+z_{\rm L})^3\sigma_{\rm SIS},
\label{dpdz_notheta}
\end{equation}
where we normalize by the total probability of strong gravitational
lensing:
\begin{equation}
 P(z_{\rm S})=\int_0^{z_{\rm S}}dz_{\rm L}P(z_{\rm L}; z_{\rm S}).
\end{equation}
From this distribution of $z_{\rm S}$, we can calculate the probability
distribution for $\Delta t(\mu_+-1)/\theta^2$:
\begin{equation}
P(\Delta t(\mu_+-1)/\theta^2)
=P(z_{\rm L}; z_{\rm S})\left(\frac{dF}{dz_{\rm L}}\right)^{-1}.
\label{dtmu2}
\end{equation}
Therefore we can predict the probability distribution for a combination
of the time delay, the magnification and the image separation. Figure
\ref{fig:predict_notheta} plots this probability distribution for
various $z_{\rm S}$. Errors in $\mu_+$ are not important for the same
reason described in \S \ref{sec:case2}.

Again we find a fitting formula for $P(\Delta t(\mu_+-1)/\theta^2)$:
\begin{eqnarray}
 P(\Delta t(\mu_+-1)/\theta^2)d(\Delta t(\mu_+-1)/\theta^2)&&\nonumber\\
&&\hspace*{-50mm}
=\frac{1}{\sqrt{2\pi}\sigma}
\exp\left[-\frac{\left\{\ln\left(\Delta t(\mu_+-1)/\theta^2\right)
-\ln a\right\}^2}{2\sigma^2}\right]\nonumber\\
&&\hspace*{-47mm}\times d\ln\left(\Delta t(\mu_+-1)/\theta^2\right),
\end{eqnarray}
\begin{equation}
 a=-110+158z_{\rm S}^{0.212}\,{\rm [day]},
\end{equation}
\begin{equation}
 \sigma=0.846.
\end{equation}
This formula is valid for $0.5\lesssim z_{\rm S}\lesssim 4.0$, and the
accuracy is $\lesssim 10$\% around the peak (within $\sim 1.5\sigma$).

\vspace{0.5cm}
\centerline{{\vbox{\epsfxsize=8.3cm\epsfbox{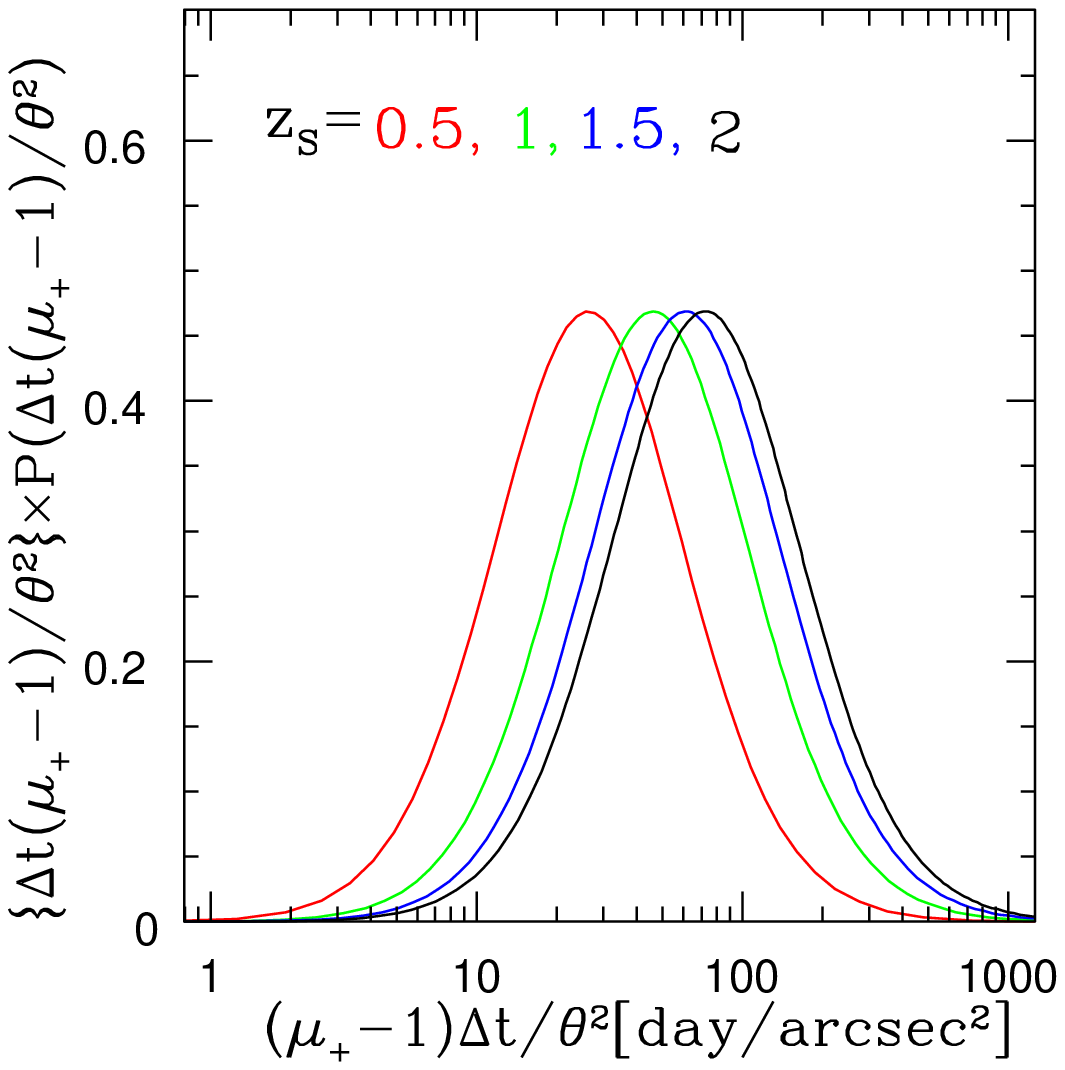}}}}
\figcaption{Predictions for the trailing image in ``Case 3'' (\S
 \ref{sec:case3}). Probability distributions of
$(\mu_+-1)\Delta t/\theta^2$ (eq. [\ref{dtmu2}] times
$(\mu_+-1)\Delta t/\theta^2$) are plotted for various $z_{\rm S}$. A
 lambda-dominated universe is assumed. \label{fig:predict_notheta}}
\vspace{0.5cm}

\subsection{Case 2 \& NFW profile}\label{sec:case2nfw}

\begin{figure*}[t]
\epsscale{1.8}
\plotone{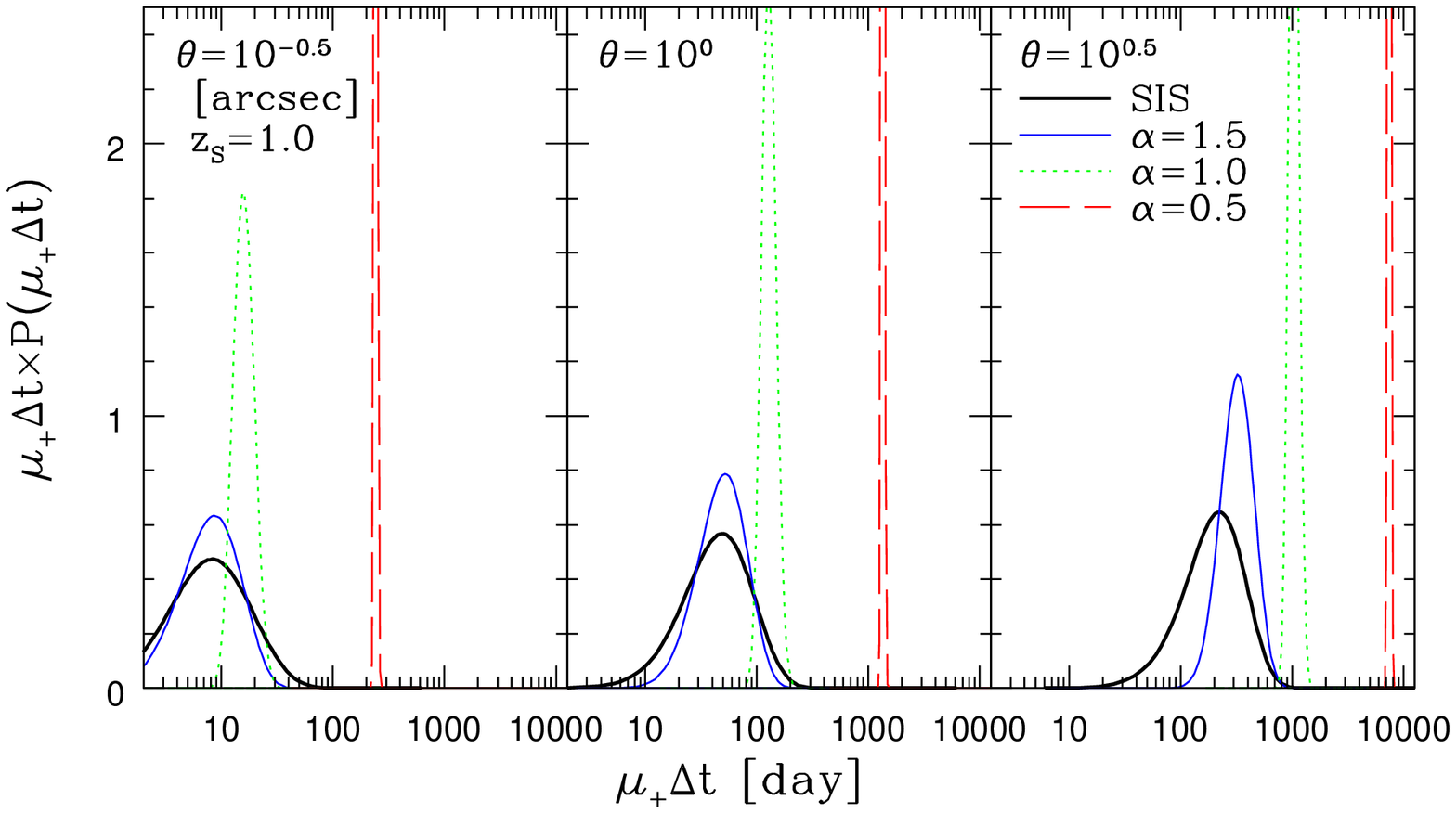}
\caption{Predictions for the trailing image in ``Case 2'' using a
 generalized NFW density profile (eq. [\ref{nfw}]). For the SIS case
 probability distributions of $(\mu_+-1)\Delta t$ instead of
 $\mu_+\Delta t$ are plotted. A lambda-dominated universe is assumed.
 \label{fig:predict_theta_nfw}} 
\end{figure*}

In this subsection, we consider a case in which the density profile of
the lensing objects is well described by the generalized NFW profile
(eq. [\ref{nfw}]) instead of by a SIS.  We retain the velocity function
of galaxies (eq. [\ref{velfunc}]) previously used, since we mainly want
to elucidate the effect of different density profiles. In this case,
differential time delays and image separations are approximated as
\citep{oguri02a}
\begin{equation}
 \Delta t=\frac{2 r_{\rm s}^2x_{\rm t}D_{\rm OS}}
{cD_{\rm OL}D_{\rm LS}}(1+z_{\rm L})y,
\end{equation}
\begin{equation}
 \theta=\frac{2 r_{\rm s}x_{\rm t}}{D_{\rm OL}},
\end{equation}
where $x_{\rm t}$ is a radius of the tangential critical curve
normalized by $\xi_0=r_{\rm s}$.  We also assume that $\mu_+$ is
approximately given as
\begin{equation}
 \mu_+=\frac{\mu_{\rm t0}}{2}\frac{y_{\rm r}}{y},
\end{equation}
where $y_{\rm r}$ is the radius of the radial caustic. The explicit form
of $\mu_{\rm t0}$ may be found in \citet{oguri02a}. From these,
\begin{equation}
 \mu_+\Delta t=\theta^2\frac{y_{\rm r}\mu_{\rm t0}}{4cx_{\rm t}}
\frac{D_{\rm OS}D_{\rm OL}}{D_{\rm LS}}(1+z_{\rm L}),
\end{equation}
and thus we can derive the probability distribution of $\mu_+\Delta t$.
Figure \ref{fig:predict_theta_nfw} shows the probability distribution of
$\mu_+\Delta t$ for inner slopes $\alpha=0.5$, 1.0 and 1.5 in dashed,
dotted and thin solid curves, respectively. This figure indicates that
a smaller $\alpha$ predicts larger values of $\mu_+\Delta t$.
The dependence of $\Delta t$ alone on $\alpha$ has the opposite sign
; a steeper inner slope of the density have larger time delays on
average \citep{oguri02a}. The reason is that a shallower density profile
tends to produce larger $\mu_+$ values and thus to cancel the dependence
of $\Delta t$ and $\mu_+$ on $\alpha$. The effect of varying the assumed
lens density profile is thus significantly reduced by competing effects.

\section{Discussion}

We have studied strong gravitational lensing of distant supernovae with
particular attention to their magnification and time-delay statistics.
Since supernovae are both ``standard candles'' at peak brightness and
transient phenomena, unlike more conventional lensed sources such as
quasars, they have some unique and relatively unfamiliar properties.

One such feature of supernova lensing is that we can determine the
magnification factor directly from observations, not just the
magnification ratios between different images. Since SNe Ia are known to
be an excellent standard candle, unusually luminous supernovae are always
strong candidates for strong lensing. We have shown that the location and
the time delay of the trailing images of a lensed supernova with a given
magnification factor can be predicted with useful accuracy. We have
further considered several cases, depending on whether the lensing
galaxy can be identified or not and whether its redshift is known or not.
Such predictions will allow targeted observing programs to study
exceptionally interesting phases of SN explosions and the determination
of extremely accurate time delays. While we have mainly considered
lenses with an SIS density profile, we find that our results are not
qualitatively sensitive to variations in this density profile, and
specifically, that generalized NFW profile lenses produce quite
similar effects.

Due to the finite duration of the event, strongly lensed SNe will not
always have multiple images observable simultaneously. In many cases,
the second (usually fainter) image will appear after the first
(brighter) image has faded away. This leads to an observational bias
against the detection of multiple lensing images in any realistic SNe
survey. We have calculated this time delay bias analytically, on the
basis of time delay probability distributions derived by
\citet{oguri02a}.  We find that time delay bias significantly changes
the expected number of lensed SNe, especially at wide separations. More
specifically, if the observational survey lasts of an order of a year,
the lensing probability is suppressed by more than one order of
magnitude at $\theta\sim 10''$. The suppression of the lensing
probability is greater for a steeper inner density profile in the
lensing objects. This is simply because a steeper inner profile yields a
larger time delay \citep{oguri02a}. We note that the time delay bias may
have less effect on ground based SN surveys which are expected to operate for a
longer term, such as LSST or a SN pencil beam survey like that proposed by
\citet{wang00}.

In this paper we have studied spherically symmetric lenses. Although the
inclusion of small ellipticities has little effect on the lensing cross
section \citep{blandford87,kochanek87}, it can yield lensing systems
with four images.  In such a more realistic case, our predictions for
the time delays and image separations should correspond approximately
to those for the pair of images with the largest separation in each
quadruple (their fractional errors are expected to be of order the
ellipticity). Time delays among the other images can be much smaller
than our predictions; e.g., the time delay between A1-A2 in PG1115+080
is expected to be much smaller (of order of one hour) than the time
delay between B-C ($\sim$ 25 days) \citep[e.g.,][]{keeton97}. The time
delay bias for such systems might then be greatly reduced. In addition,
four images usually appear when the position of the source is close to
that of the center of the lens galaxy.  This will also cause four-images
systems to have systematically smaller time delays. We thus expect that
the ratio of four image to two image lens systems will be larger for
lensed SNe than for other lens systems in which all images are
continuously present.

Another pleasing aspect of strongly lensed SNe Ia is that they allow one
to make a fairly straightforward {\it theoretical} prediction for a
cosmologically distant phenomenon that is then subject to direct
quantitative verification on a humanly practical time scale.  This
possibility is relatively rare in astronomy except in case involving
intrinsically periodic phenomena, such as orbits or pulsar emission,
where the ``prediction'' is basically empirical extrapolation rather than
truly theoretical.  It is particularly rare in a cosmological context.
Although one certainly would not expect major surprises in comparing
such predictions to future observations, it is still an important
opportunity to test and validate our basic understanding of cosmology
and gravitational theory.

\acknowledgments
This research was supported in part by the Grant-in-Aid for Scientific
Research of JSPS (12640231, 14102004) and by NASA grant NAG5-9274.

\end{document}